\documentclass[fleqn,10pt]{wlscirep}
\usepackage[utf8]{inputenc}
\usepackage[T1]{fontenc}
\usepackage{lineno}
\title{Optical chiral microrobot for out-of-plane rotation}

\author[1]{Alaa M. Ali}
\author[2,3]{Edison Gerena}
\author[1]{Julio Andrés Iglesias Martínez}
\author[1]{Gwenn Ulliac}
\author[1]{Brahim Lemkalli}
\author[1]{Abdenbi Mohand-Ousaid}
\author[2]{Sinan Haliyo}
\author[1]{Aude Bolopion}
\author[1,*]{Muamer Kadic}
\affil[1]{Université Marie et Louis Pasteur, CNRS, Institut FEMTO-ST (UMR 6174), Besançon, 25000, France}
\affil[2]{Sorbonne Université, CNRS, Institut des Systèmes Intelligents et de Robotique (UMR 7222 ISIR), Paris, France.}
\affil[3]{MovaLife Microrobotics, Paris, France}

\affil[*]{muamer.kadic@femto-st.fr}


\begin{abstract}
Optical microrobots (OPTOBOTs) have garnered significant interest, particularly in the medical field, due to their potential for precise cell manipulation in various biological studies and microsurgical applications. Previously described OPTOBOTs demonstrate multiple degrees of freedom, yet improvements are needed, especially in achieving reliable out-of-plane rotation. Here, we propose an OPTOBOT design based on chirality that enables full-cycle out-of-plane rotations using optical tweezers. The OPTOBOT has an arrow-like structure with two handles aligned on the same axis, maintaining its horizontal orientation and facilitating controlled movement. Additionally, the OPTOBOT’s tail is a chiral helix, which induces repetitive out-of-plane rotations around its longer axis when targeted by a laser beam that is due to broken axial parity. Finite element analysis is employed to design the OPTOBOT and assess its capacity to generate mono-directional high optical torque. Experimental results confirm various actuation modes, supporting future integration of OPTOBOTs in complex micromanipulation tasks.

\end{abstract}
\begin{document}

\flushbottom
\maketitle
\thispagestyle{empty}

\section*{Introduction}

Microscopic objects manipulation has advanced with the development of lasers, which concentrate light power into small areas, enabling the use of optical forces for precise control. Generally, light applies force on microparticles in which radiating pressure, or the force exerted on an object per unit area, is proportional to the rate of change of linear momentum, leading to pushing particles in the same direction of light propagation. However, in the case of optical tweezers, the use of a highly focused laser beam creates a gradient of light intensity inducing forces on tiny particles near the focal point of the beam enabling their trapping \cite{Volpe2023}. 
In essence, optical tweezers can be used to successfully handle, rotate, and move microobjects in specific directions. The advancements in optical tweezers led to great discoveries and applications in different fields such as nanotechnology and nanofabrication \cite{PMID:24202536, dc758862275440ff8b4a2db2f20b7c19,Mcleod2009ArraybasedON, Mcleod2008SubwavelengthDN}, spectroscopy \cite{doi:10.1021/acs.accounts.0c00407} quantum science \cite{12661,refId0}, plasmonics \cite{PMID:33731693}, and particularly biology. 
 For instance, optical tweezers can be used in trapping a single microorganism like a virus \cite{PMID:2250707,PMID:30872888} and bacterium \cite{PMID:2250707, ZhangZheng2019Mrbw}. Also, in the context of cell studies and manipulations, automated optical tweezers systems were developed to control cell transport \cite{tmech2017}, rotation \cite{cellrotation2015}, and mechanical studies \cite{tmech2011}. However, direct control of the cells with the laser can cause damage to them. Thus, optical microrobots (OPTOBOTs) are being developed as means of controlling and manipulating cells.

High-precision control of microrobots with optical tweezers is achieved by applying an optical trap to specific points, which generates multiple degrees of freedom for actuation \cite{xu2024power}. This has led to the development of various applications to manipulate microparticles and cells. For instance, Ta et al. \cite{Tmech2020} used optically-driven microbeads to act as fingertips to grasp microobjects simultaneously. Also, the grippers can have different designs that can be fabricated by the two-photon lithography technique to be able to grasp different objects with different shapes \cite{microhand2016}. Moreover, cell transport is conducted \cite{celltransport2019}, and also multiple degrees of freedom can be achieved \cite{xu2024gradient}. For instance, Gerena et al. \cite{Edison} introduced teleoperated OPTOBOTs controlled with multiple traps giving $6$ degrees of freedom control, and according to the design of the end-effector of the microrobot, single-cell or group of cells can be transported with dextrose control. Other functions that can be conducted with OPTOBOTs in the medical and biological fields include cell puncture \cite{puncture}. However, the puncture technique introduced before depends on the thermal effect, which may cause harm to the cells. The presented previous work could achieve different functions and movements in multiple degrees of freedom, but the out-of-plane movement was limited.

Out-of-plane rotation of OPTOBOTs is demanded especially for medical applications, as it is needed for cell drilling and handling in specific orientations for some microsurgeries and studies such as single-cell analysis, sperm injection, nuclear transplantation, and multi-dimensional imaging \cite{Hu:19,10.1002/adom.202000543,avci2017laser,articlewheel}. Furthermore, it is also important for the control of microfluidic chips to form pumps and micro-mixers. To achieve out-of-plane rotation, a holographic optical tweezer can be used \cite{Hu:19}. It relies on several focused light beams, each controllable individually. This allows for out-of-plane rotation by independently adjusting the depth of the focal point for various laser traps. Nevertheless, the holographic setup is complex, bulky, and costly \cite{10.1002/adom.202000543}. Therefore, other simpler planar optical tweezer systems based on time-multiplexing techniques may be more favorable to be used. Time-multiplexing techniques depend on deflecting the laser beam between several trap positions with high frequency. This is achieved by using actuated mirrors, acousto-optic deflectors, or electro-optic deflectors. This allows the trapping of multiple objects or multiple parts of the microrobot and due to the high frequency of the trap position rotation, trapped points do not have time to be lost from the trap. However, these techniques are often limited to the planar motion of the traps \cite{gerena2019high}. Therefore, different designs of OPTOBOTs were introduced in the literature to give out-of-plane rotation using planar time-multiplexing optical tweezers. There are different approaches used to give out-of-plane rotation. The first approach depends on sequential trapping of a curved structure and by changing the distance between two traps, the orientation of the structure changes, and because it is curved this will induce out-of-plane rotation \cite{10.1002/adom.202000543}. Another approach is using a structure with multi-components connected by joints. Trapping the base to fix it and moving another trap on the mobile head; leads to its rotation due to the joint \cite{avci2017laser,micromechanisms2016}, but some optimizations are needed to avoid adhesion between the joint components. The third approach depends on the scattering force to push a paddle-wheel-like structure \cite{articlewheel}. However, this design suffers from uneven geometry making the paddle-wheel prone to tilting and difficult to control, requiring adjustments to improve stability and reduce resistance. Also, the obtained optical torque was low. The optical torque is always reported as a normalized value, which represents the ratio of the calculated torque to the maximum theoretically achievable torque under given conditions, providing a dimensionless measure of the efficiency or effectiveness of torque generation in optical systems. The maximum theoretically achievable torque ($\tau_{\rm max}$) is the highest possible torque that can be generated by an optical system, determined by the interplay of factors like optical power, the refractive index of the medium, and the physical dimensions of the object being acted upon. It represents the upper limit of torque efficiency that could be attained under ideal conditions \cite{articleturbine,RahimzadeganPhysRevB2017}, and it is calculated according to equation (\ref{eq03}).
\begin{equation}\label{eq03}
\centering
\tau_{\rm max}=P\cdot n_{\rm m}\cdot r/c,
\end{equation}
where $P$ is the laser power, $n_{\rm m}$ is the refractive index of the medium around, $r$ is the radius of the part of the helix interacting with the laser beam, and $c$ is the light velocity. The normalized optical torque in the wheel-like design was estimated \cite{articleturbine} to be limited to $9\%$. Generally, the previously proposed designs are limited to either rotation without achieving a full cycle or low optical torque.

This paper proposes a different approach based on chirality to achieve out-of-plane rotation with a high optical torque. Chirality refers to the asymmetrical geometric arrangement of components within a structure, resulting in a non-centrosymmetric configuration such as springs, which can wind in either a clockwise (right-handed) or counterclockwise (left-handed) direction. Chiral geometries in fluid mechanics can induce rotational motion or torque transmission when subjected to translational velocity, particularly in microscale systems dominated by viscous forces enabling applications such as microfluidic mixing, propulsion, and turbulence control \cite{collins2017chiral}. \textcolor{black}{Moreover, chiral structures are used in the synthesis of microswimmers that are actuated by a magnetic field \cite{magneticRobotics}}. Recently, several studies have demonstrated that the optical force direction can be altered using asymmetrical and metastructures \cite{arslan2022toward}, this can produce effects like lateral force \cite{nan2023creating} and rotation. Moreover, chiral structures were introduced before to induce rotations due to optical forces by controlling the polarization of the incident light beam. However, the rotations were in-plane \cite{RahimzadeganPhysRevB.2016, articleturbine,6989315, ChiralNGScience}. In our case, we numerically and experimentally characterize our microrobot design that has an arrow-like structure where its head is used to control the position and a chiral tail that can generate out-of-plane rotation as a result of the interaction with a laser beam, as shown in \autoref{design}(a). This rotation is achieved by breaking the axial parity by applying an optical off-axis force to the chiral part, resulting in a repetitive out-of-plane rotation of our OPTOBOT. \textcolor{black}{The main advantage of this design is that it provides continuous out-of-plane rotation with high optical torque making it promising to be implemented in microrobots with different end-effectors to help in different applications such as cell manipulation, rotation, and drilling.}  
\textcolor{black}{Table 1 summarizes the previously proposed optical microrobots  that give out-of-plane rotation compared to our proposed design}

\section*{Results and Discussions}
\subsection*{Design and modelling of the OPTOBOT}\label{sec2}
\begin{figure*}[h!] 
    \centering
  \includegraphics[width=18cm]{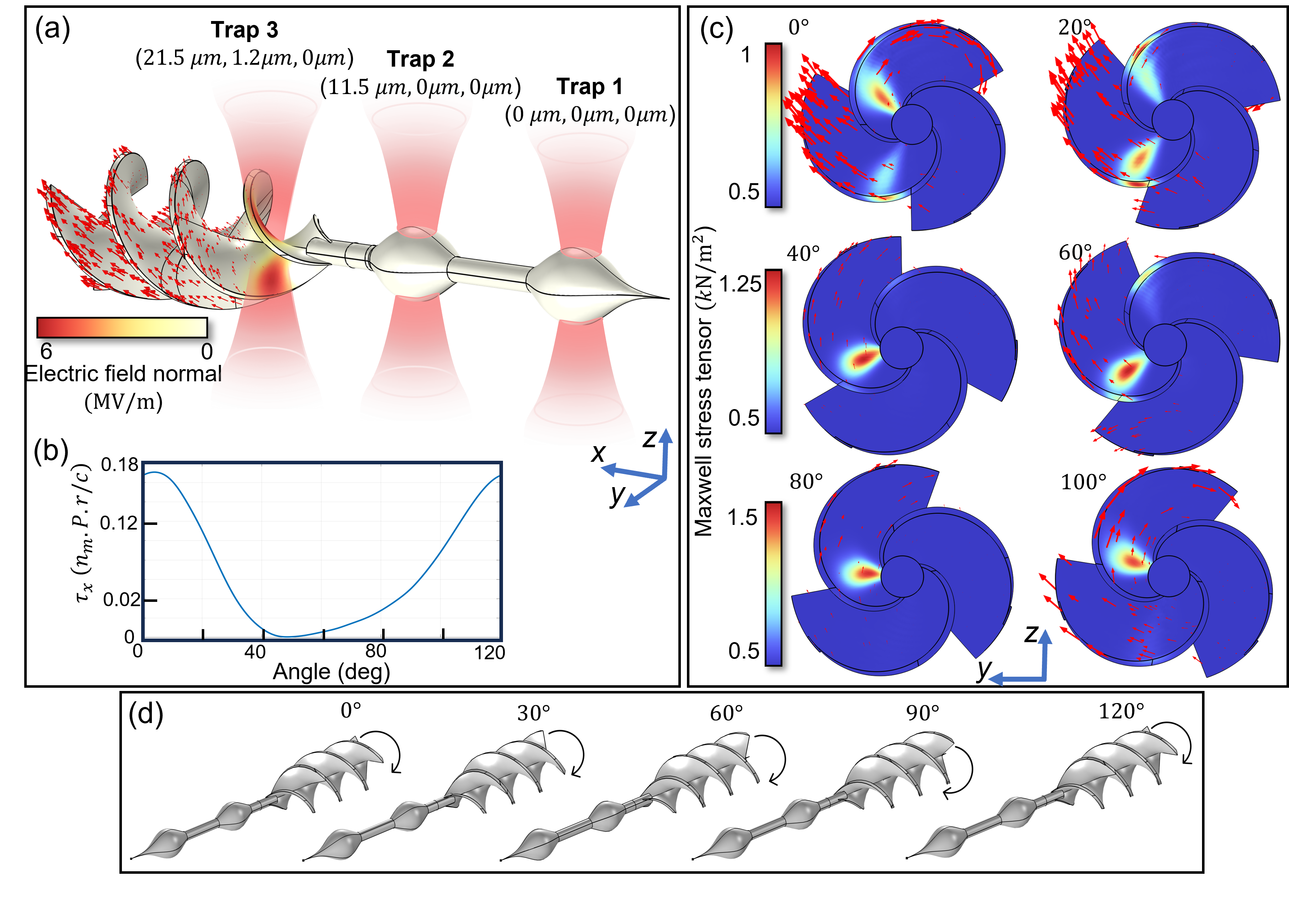}
    \caption{Principle of the optical microrobot (OPTOBOT) for out-of-plane rotation. (a) Design of the OPTOBOT showing the laser beams positions with their coordinates (Traps 1 to 3) onto it with propagation along the $z$-direction. Here, Trap 1 and 2 are used to hold the OPTOBOT while the off-axis Trap 3 is used to create pressure on the OPTOBOT's helix and generate its rotation. We depict on the helix the norm of the electric field created by the Gaussian illumination via the false color-scale on its surface. (b) The rotation of the OPTOBOT is only possible under a constant sign of the torque along the $x$-direction and we depict the normalized $\tau_x$ versus different angles occupied during a full revolution of the OPTOBOT, from $0$ to $120$ degrees (equivalent to the full rotation due to the $C_3$ rotational symmetry of the structure). Where $n_{\rm m}$ is the refractive index of the medium (water), $P$ is the laser power, $r$ is the radius of the helix part interacting with the laser beam and $c$ is the light velocity. This calculation is done while assuming the two other traps generate a stable stiffness and do not allow the OPTOBOT to move and give the only possible motion a revolution along the $x-$ axis. A positive value of $\tau_x$ is observed. In (c) we show the magnitude of the Maxwell stress tensor for different angles during the rotation. The color map is shown for each two angles, while the red arrows show the displacement field direction. The calculations were done assuming input laser power of $630 {\rm mW}$. \textcolor{black}{(d) The rotation of the robot is illustrated for different angles, the rotation is clockwise as shown by the arrow, and the OPTOBOT returns to the same configuration at angle $120^\circ$.} }
    \label{design}
\end{figure*}

\noindent To design an efficient rotating microrobot, it must be considered that the OPTOBOT is mainly controlled by optical tweezers and observed under a microscope; therefore, all the components must be on the same plane. For instance, if a cell needs to be manipulated, both the OPTOBOT and the cell must be within the same plane to be observable under a single focal plane of the microscope. Consequently, the rotation of the microrobot must occur in an out-of-plane direction to manipulate something beside it. To increase dexterity, it should enable full-cycle rotations. Additionally, the rotation process must be on-demand; thus, controlling the microrobot's translational motion and position must be separate from controlling its rotation. It is also preferable to be able to rotate the microrobot while it is either fixed or moving forward providing greater flexibility in its functions. Therefore, to enable trapping and translation motion, our proposed OPTOBOT, as depicted in \autoref{design}(a), has two optical handles (Trap $1$ and Trap $2$). These two optical handles are on the same axis to keep the OPTOBOT in place. To enable rotations, the tail is a helix with a chiral structure. When the laser beam is directed at this part (Trap $3$), the structure rotates around the $x$-axis (\autoref{design}(a)). Thus, the OPTOBOT only rotates when the third laser beam is activated. The front handle has a tip that might be used later for puncture purposes. Also, the diameter of the whole robot does not vary much around the axis of rotation and the optical handles are ellipsoidal to help in reducing the drag. It should be noted that Trap $3$ is off-axis compared to Trap $1$ and Trap $2$ and does not point to the center of the robot axis but is offset by about $1.2 {\rm \mu m}$. \textcolor{black}{This offset position was the one giving the highest torque and the most stable rotation compared to the other positions}. \textcolor{black}{A detailed geometry and the dimensions of the OPTOBT are shown in Supplementary Note 1 and figure S1.}

The optomechanical behavior of this design was studied by finite element method using COMSOL Multiphysics software (for more details, see Methods section). A linearly-polarized Gaussian beam with a wavelength of $1070 {\rm nm}$ propagating in the $z$-direction was directed to the chiral part. Then, MST was implemented in the structural mechanics module to calculate the generated optical force and torque on the microrobot (see Methods Section). Our calculations assume that the two traps on the optical handle generate a stable stiffness and prevent any motion except the rotation along the $x$-axis.\textcolor{black}{That is because the optical trap allows the rotation of the optical handles as they are symmetric around the $x$-axis, so this rotation will not affect the interaction of the light with the optical handle. However, any other movement will be unsymmetrical and will be resisted by the gradient force of the optical traps on the handles.}

The simulation results show that the electric field distribution was tailored by the chiral part leading to breaking the axial parity as it interacts with the chiral geometry (\autoref{design}(a)). \textcolor{black}{Breaking axial parity refers to the asymmetry in the interaction of light with the chiral helix, where the optical force does not behave identically under mirror inversion or rotation around the longer axis of the OPTOBOT. This violation can help in phenomena like optical torque or the transfer of angular momentum, where the behavior of light-matter interaction becomes directionally dependent.}
MST was calculated for the helix with different angles of rotation from $0$ to $120$ degrees around the $x$-axis as its shape exhibits rotational symmetry of order $3$, as depicted in \autoref{design}(d). It is observed that the stress distribution within the structure is asymmetric and it varies according to the angle (\autoref{design}(c)). Consequently, a torque is generated, and its value also changes according to the angle of rotation. However, the initiated rotational motion is always in the same direction as the helix's chirality. As shown in \autoref{design}(b) representing the calculated normalized torque for different angles of rotation mentioned in Supplementary Data 1, the generated torque is always positive, and the OPTOBOT should not exhibit any reverse rotations. Furthermore, the variation of the torque with the angle can also be attributed to the contribution of the scattering force to different positions. When a part of the helix is in direct contact with the focal point, the laser power is higher and hence the optical force and torque increase.  The normalized optical torque is calculated by referencing the maximum achievable optical torque $\tau_{max}$ (equation \ref{eq03}). Using this normalization for a standardized evaluation of the optical torque's efficiency \cite{articleturbine}, facilitates comparisons across different experimental setups and conditions. The maximum calculated normalized torque is $0.17$ ($17\%$ of $\tau_{max}$) which is, to our knowledge, the highest reported value for the out-of-plane torque in optical microrobots.

\subsection*{OPTOBOT fabrication}

\begin{figure*}[!h] 
    \centering
     \includegraphics[width=18cm]{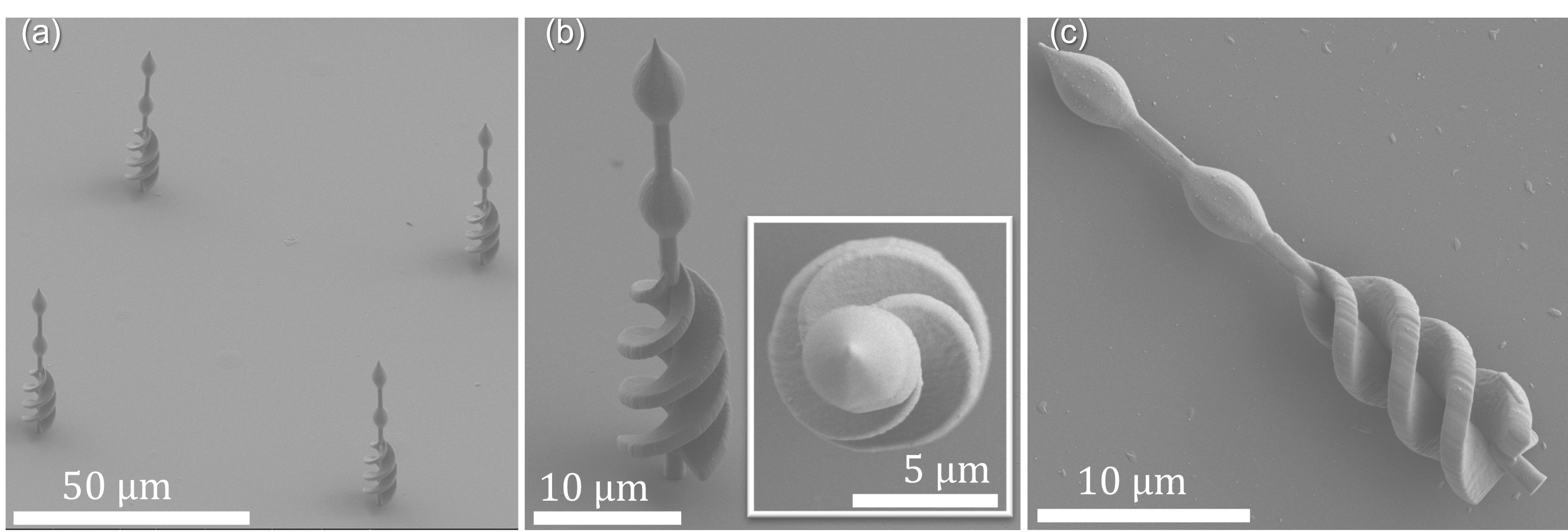}
    \caption{SEM images of the optical microrobot fabricated using the two-photon lithography printing technique. The robots were printed to be standing vertically over the pillar connected to the substrates. (a) We show an array of robots standing perpendicularly to the substrate, (b) a top and tilted view with an angle of $45$ degrees of a standing robot, and (c) a top-view of a robot detached from the substrate to show clearly the printing of the different components of the OPTOBOT; the helix, optical handles, and the indentation at the front optical handle.}
    \label{sem}
\end{figure*}

 The robots were printed vertically to decrease the adhesion to the substrate, and a thin pillar with a $660{\rm nm}$ diameter was printed under the helix to connect the OPTOBOTs to the substrate. Therefore, all the robots were supposed to stand vertically over the pillar attached to the substrates.
 The robots were printed vertically to decrease the adhesion to the substrate, and a thin pillar with a $660{\rm nm}$ diameter was printed under the helix to connect the OPTOBOTs to the substrate. Therefore, all the robots were supposed to stand vertically over the pillar attached to the substrates.
 The OPTOBOTs depicted in \autoref{sem} were fabricated by a $3$D direct laser writing printer based on the two-photon absorption lithography technique, which is a high-resolution $3$D printing process that utilizes a femtosecond laser to induce polymerization at the focal point within a photosensitive resin. This technique relies on the simultaneous absorption of two photons, which provides the energy necessary to initiate polymerization only at the precise focal spot. As a result, this technique enables the creation of intricate microstructures with sub-micrometer resolution. By scanning the laser focal point in three dimensions, complex $3$D micro- and nano-structures can be constructed layer by layer, making it an ideal method for fabricating sophisticated microrobots. \textcolor{black}{Different resins can be used in two-photon lithography; however, they vary in their resolution and properties. We chose IP-L $780$ resin for our printing as it provides high resolution, where the smallest feature that can be printed is $200 \, {\rm nm}$. Moreover, it is printed in the oil immersion configuration on substrates used directly in the optical setups. Therefore, there is no need to transfer the OPTOBOTs from one substrate to another. For better optical manipulation, the used material has to be of high refractive index, IP-L has a refractive index of 1.5, and the other resins normally used for microrobots fabrication have refractive indices in the same range or even lower. Therefore, we chose the resin that gives us higher resolution and is more practical in the experiments.} 
 The robots were printed vertically to decrease the adhesion to the substrate, and a thin pillar with a $660{\rm nm}$ diameter was printed under the helix to connect the OPTOBOTs to the substrate. Therefore, all the robots were supposed to stand vertically over the pillar attached to the substrates.
The SEM images given in \autoref{sem} demonstrate different robots with a side view angle of $45$ degrees to show the robots standing on the substrates, and a zoomed picture to show the whole body including its different components: (i) the helix that was printed with high resolution and without any adhesion between its parts, (ii) the pillar and the optical handles where the front one has a thin indentation to help to interact with cells; this part was also printed with high resolution; although its diameter is around $200 {\rm nm}$. We also depict the top view of a standing robot that only shows the head and part of the helix, and a top view of another robot with its pillar detached from the substrate to show its structure more clearly. The OPTOBOTs were then detached from the substrate using a three-axis micromanipulator system (Supplementary Movie 1) to be ready for testing by the optical tweezers system.
\begin{figure*}[!h] 
    \centering
     \includegraphics[width=18cm]{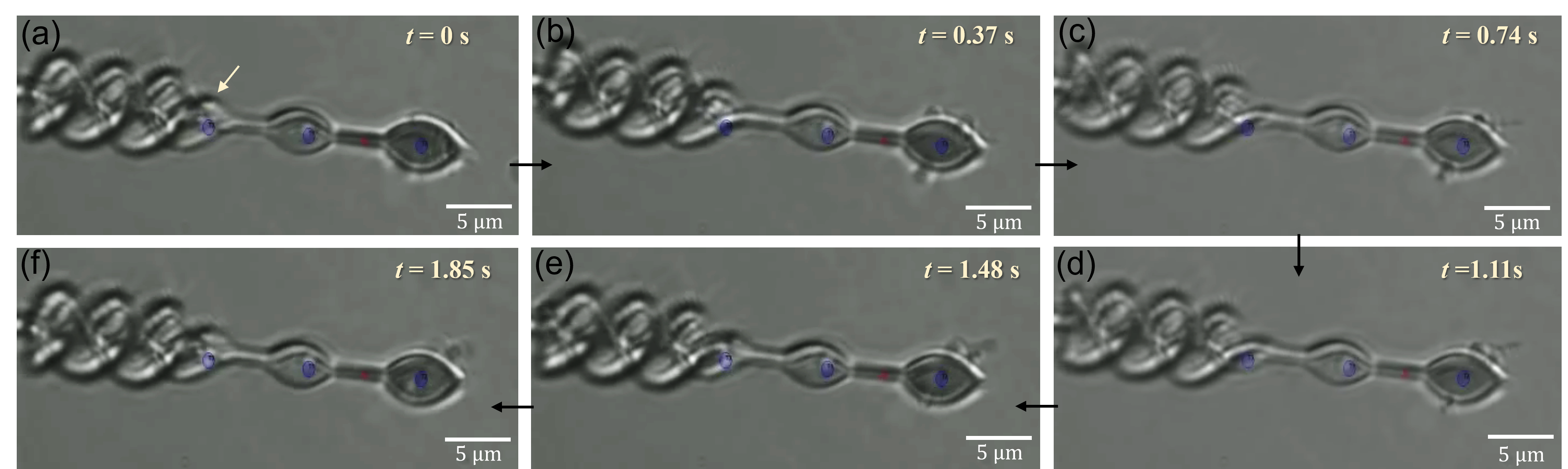}
    \caption{Optical microscope images of the optical microrobot during one-third of the rotational cycle. Snapshots are taken from Supplementary Movie 2 at different times during the rotation of the OPTOBOT (a-f) when the third trap is focused on the chiral part while the two other traps are used for keeping the robot in place. The laser power used here is $6.3\%$. To track the rotation, the first turn of the helix (indicated by a yellow arrow in the first picture) was followed. The robot gets a similar configuration after $1.85 {\rm s}$ (panels (a) and (f)), which indicates a third of a rotational cycle due to the $C_3$ symmetry of the robot.}
    \label{opticalimages}
\end{figure*}
\subsection*{Experimental demonstrations}
An Optical Tweezers setup detailed in Supplementary Note 4 was used to actuate our OPTOBOT. In our experiment, we show two aspects: the influence of the laser power on the rotational capability and the overall motion capabilities of the OPTOBOT, especially its ability to produce full out-of-plane rotations. 

For that, the OPTOBOT is actuated by three laser beams (one beam is time-shared between several trapping positions): two for trapping the optical handles, and the third one for rotating the chiral helix. The process begins with activating the first two traps to hold the micro-robot and control its position; then, the third trap is activated to start the rotation. 

\subsubsection*{Influence of the laser power}
The laser source used in our experiments has a maximum output power of $10 {\rm W}$. During this experiment, the percentage of the power used in the manipulation was tuned to analyze its influence on the OPTOBOT. Here, we are demonstrating the effect of four different power values: a low power of $6.3\%$, moderate power of $15\%$, a high power of $30\%$, and a maximum power of $50\%$, to show the limitations of each power and the processes that can be done. 

The OPTOBOT shown in \autoref{sem} was tested using laser power of $6.3\%$ to apply an optical force and trap the microrobots. The OPTOBOT was trapped and an optical force was added successfully to the third trap to induce rotation. This led to the rotation of the OPTOBOT around its longer axis in the out-of-plane direction with an angular velocity of $0.18 {\rm Hz}$ (\autoref{opticalimages}, Supplementary Movie 2). The rotation was repetitive as long as the third trap was active and the OPTOBOT did not show any reverse rotation, as expected by our computational model. By adding more laser power, the rotational velocity can be increased up to $0.32  {\rm Hz}$ (Figure S4); however, there is a limitation on increasing the power to higher values. 

\begin{figure*}[!h] 
    \centering
     \includegraphics[width=18cm]{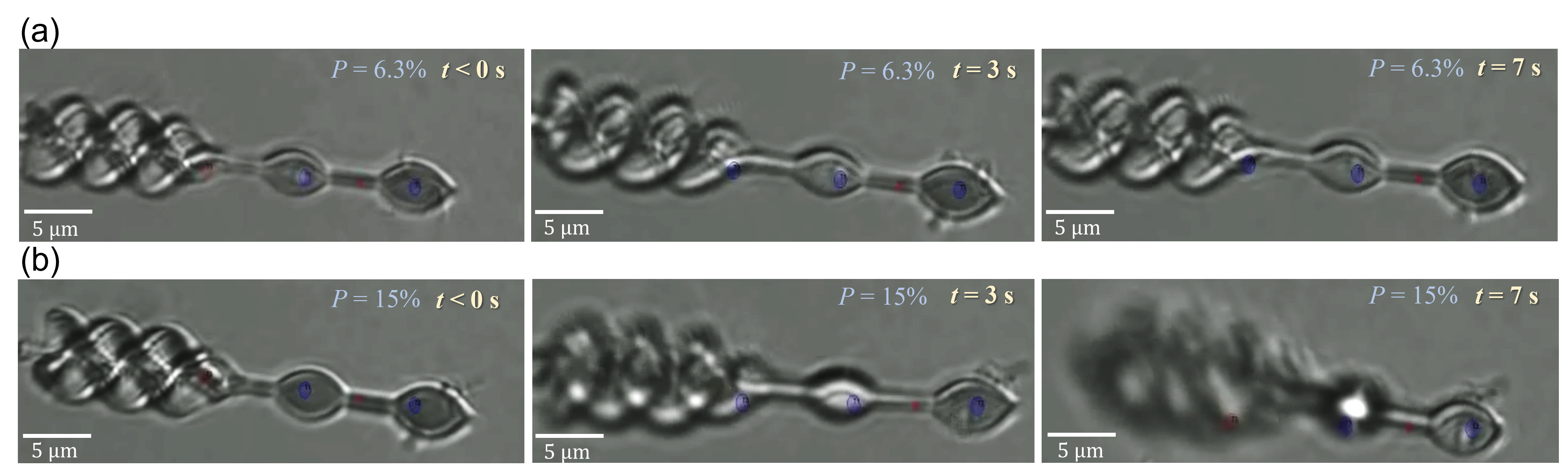}
    \caption{Effect of increasing the laser power from $ 6.3\% $ (row a) to $ 15\% $ (row b). When using only two traps on the optical handles the structure is stably trapped and is totally horizontal. In the case of the low power, after activating the third trap on the helix, the helix starts to rotate and it is able to stably rotate as long as the third trap is activated. However, in the case of the higher power, the helix rotates faster but at the same time it seems to be blurred, which means that it is becoming out of focus as it is pushed by the laser force, and after $7$ seconds it becomes completely lost from the traps (Supplementary Movie 3).}
    \label{lp}
\end{figure*}

In \autoref{lp}, a comparison is conducted between two laser power sets, a low power of $6.3\%$ and a moderate power of $15\%$. The results show that, with the lower power, the OPTOBOT rotates stably as long as the third trap is activated. In contrast, at higher power, the robot is stable only when the traps on the two optical handles are active. However, when the third laser beam is activated, the robot rotates, but it is simultaneously pushed in the direction of the laser propagation leading to its separation from the trap (Supplementary Movie 3).  By increasing the power, when the laser interacts with the chiral helix, the rotation speed increases. However, the laser beam has a scattering (pushing) force, and by increasing the power, this pushing force increases, leading to pushing the whole robot and its separation even from the optical traps holding the optical handles. \textcolor{black}{Therefore, it is necessary to find the power that achieves the balance to be able to trap the optical handles and at the same time does not cause pushing to the whole robot. More optimization to the OPTOBOT design may also lead to better trapping without losing it in higher powers.} 

\begin{figure*}[!h] 
    \centering
     \includegraphics[width=18cm]{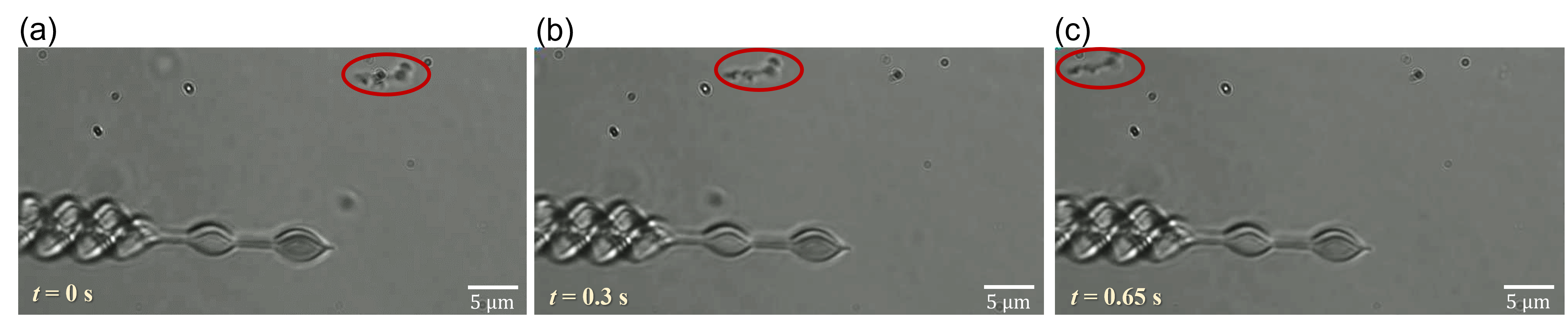}
    \caption{Using laser power of $30\%$ to achieve high translational velocity. (a) The OPTOBOT at the beginning of the translation. (b) The OPTOBOT after $0.3 {\rm s}$. (c) The OPTOBOT after moving $34 {\rm \mu m}$ in $0.65 {\rm s}$ giving a translational velocity of $52 {\rm \mu m/s}$. The translation can be followed relative to the particle with the red circle around it. The snapshots are taken from Supplementary Movie 4.}
    \label{velocity}
\end{figure*}

Despite this limitation, using high power can be of interest for just trapping the two optical handles, to give an extra function of the translation with high velocity. Indeed, the higher power increases the trap stiffness, so the OPTOBOT is more firmly held. When translating with higher velocity the drag force increases; therefore, higher trap stiffness is needed to overcome the drag force. In \autoref{velocity}, it is shown that the OPTOBOT can move with a velocity of up to $52 {\rm \mu m/s}$ with power $30\%$ (Supplementary Movie 4). However, there is still a limitation on adding more power even on the trapping handles only, because the laser trap is propagating as a Gaussian beam. Thus, even if the trap is directed to the optical handle; part of the Gaussian beam interacts with the chiral helix, and by increasing the power; the intensity of this part interacting with the chiral part increases; hence its effect is pronounced. In supplementary Supplementary Movie 5, it is shown that by just activating the two traps on the optical handles; the helix starts to rotate and push at the same time, leading to the loss of the whole OPTOBOT after just $3$ seconds. Therefore, choosing the suitable power for either rotation or translation of the OPTOBOT is crucial for achieving effective performance.

\textcolor{black}{The size limitation was also studied and shown in Supplementary Note 3.}The same effect of the Gaussian beam interactions can be observed if a smaller OPTOBOT is used where the optical handles and the helix are near to each other ( Supplementary Movie 6).  From the conducted experiments investigating the laser power and the size limitations; it can be concluded that when designing an OPTOBOT the optical handles should have enough distances between each other and between the helix. Thus, the Gaussian laser beam does not cause any unintended indirect interaction. However, it is always a trade-off between the most suitable distances and the fabrication limits. For example, in our case increasing the distance between the optical handles in the helix will lead to a robot with a high aspect ratio, which may cause its bending during the fabrication or the manipulation. 
\\
\textcolor{black}{Moreover, this depends also on the type of optical tweezers device used as the Gaussian beam properties would differ. For example, we conducted experiments using other commercial optical tweezers device (Bruker JPK NanoTracker™ 2) (see Supplementary Note 4), and we could reach a higher range of power and the velocity of rotation could reach up to $1.15\rm{Hz}$ (Figure S4).}

\subsubsection*{Demonstration of the different capabilities of the OPTOBOT}\label{sec6}
By using laser power of  $ 6.3\% $, a series of experiments were conducted to demonstrate the motion capabilities of the OPTOBOT. The OPTOBOT can achieve a full out-of-plane rotation as long as the third trap on the helix is activated. Therefore, the rotation can be turned on or off according to the activation or deactivation of the third trap (Supplementary Movie 7). In addition, the rotation can be maintained while the OPTOBOT is fixed or moving (Supplementary Movie 8), and as shown before, the robot can be translated without rotation with high velocities (Supplementary Movie 4). The ability to control the OPTOBOT with different options and capabilities is highly advantageous for various applications. The precise manipulation of the OPTOBOT's rotation and movement allows for an efficient and targeted process, especially in biological cell handling. The capability to maintain rotation while the OPTOBOT is either fixed or in motion adds an extra dimension of functionality, ensuring consistent performance in dynamic environments. The ability to translate with high velocity makes the OPTOBOT highly responsive and can achieve different tasks faster. Moreover, the OPTOBOT can be actuated using different optical tweezers system and by adjusting the conditions of the laser beam, the velocity of rotation can reach up to 1.15 Hz (Supplementary Movie 9 and Figure S4); which -to our knowledge- considered high velocity compared to the other optical microrobots previously presented in the literature. This controlled and adaptable system paves the way for advanced operations where speed, precision, and flexibility are required.

\section*{Conclusion}\label{sec7}
In this article, we employed chiral broken axial parity to achieve full-cycle out-of-plane rotation of a microrobot actuated by optical tweezers. To accomplish this rotation, we used three focal regions on the OPTOBOT: two traps to control its position and keep it horizontally stable, and one focal point on the chiral helix to induce rotation. The OPTOBOT, designed with the finite element method, gives a normalized optical torque of $17\%$, which is a high value compared to the other designs giving out-of-plane rotation.
Moreover, The OPTOBOT demonstrated highly satisfactory performance during experimental optical testing matching our theoretical computations. The proposed approach of using chirality to produce full-cycle out-of-plane rotation, given for the first time in this context to our knowledge, opens the way for higher rotational velocities by optimizing the chiral helix design. The effect of the OPTOBOT size and the laser power were studied; those studies can help in understanding the limitations and guiding to improve the OPTOBOT design to achieve better performance. Furthermore, we demonstrated different capabilities of the OPTOBOT as it gives different choices of actuation: on-demand rotation with a rotational velocity of $0.18-\textcolor{black}{1.15} {\rm Hz}$, translation with speed up to $ 52 {\rm \mu m/s}$, or combined rotation and translation for more complex processes. The rotational velocity achieved here is considered high as compared to the other reported out-of-plane rotating optical microrobots.  Therefore, this technique can help in the development of dexterous microrobots to be used in biological cell handling and microsurgery procedures.
\textcolor{black}{Future work can include optimizing the structure to achieve better results and conducting real applications in the field of microsurgeries.}
\section*{Methods}
\subsubsection*{Numerical simulations}
\noindent The optomechanical behavior of this design was studied by finite element method using the commercial software COMSOL Multiphysics. Firstly, using the Radio frequency module, a linearly-polarized Gaussian beam with a wavelength of $1070 {\rm nm}$ propagating in a water medium in the $z$-direction was directed to the chiral part of the OPTOBOT with refractive index $1.5$. To calculate the electric field distribution on the helix and the electromagnetic Maxwell Stress Tensor (MST).
MST is a fundamental concept in electromagnetism that describes the flow of electromagnetic momentum through space. It encapsulates the effects of electric and magnetic field components of an electromagnetic wave giving mechanical stress on the surface interacting with the incident light. The tensor is particularly useful in calculating the force and torque on objects due to electromagnetic fields. MST, denoted as $\mathbf{T}$, is a second-rank tensor given by:
\begin{equation}\label{eq01}
T_{ij} = \varepsilon_0 \left( E_i E_j - \frac{1}{2} \delta_{ij} \mathbf{E}^2 \right) + \frac{1}{\mu_0} \left( B_i B_j - \frac{1}{2} \delta_{ij} \mathbf{B}^2 \right),
\end{equation}
where $\varepsilon_0$ is the permittivity of free space, $\mu_0$ is the permeability of free space, $E_i$ and $E_j$ are the components of the electric field vector $\mathbf{E}$, $B_i$ and $B_j$ are the components of the magnetic field vector $\mathbf{B}$, and $\delta_{ij}$ is the Kronecker delta, which is $1$ if $i = j$ and $0$ otherwise.

MST allows us to compute the electromagnetic force $\mathbf{F}$ on a particle by integrating the tensor over the surface $\mathbf{S}$, and accordingly, the torque, $\boldsymbol{\tau}$, can be calculated by the vector product between the force and the position vector $\mathbf{R}$.
\begin{equation}\label{eq02}
\begin{array}{ccc}
   \mathbf{F}  & =  & \int_S \mathbf{T} \cdot d\mathbf{S},  \\
   \boldsymbol{\tau}  &=  &\mathbf{R} \times \mathbf{F},  
\end{array}
\end{equation}
where $d\mathbf{S}$ is the differential area vector on the surface. The MST provides a comprehensive framework for understanding the interaction between electromagnetic fields and mechanical forces by describing how electromagnetic fields exert force on an object. This tensor describes the relationship between the electric and magnetic fields and the resulting force distribution within a region, allowing for precise analysis of how electromagnetic fields can generate mechanical stresses and forces on optically manipulated structures.

Following this, we leveraged the structural mechanics module to determine the displacement of the chiral component of the OPTOBOT. For this analysis, our calculations assume that the two traps on the optical handle generate a stable stiffness and prevent any motion except the rotation along the $x$-axis. This assumption simplifies the model and focuses on the rotational dynamics of the chiral part, which is crucial for understanding its behavior under the influence of the Gaussian beam. The calculated torque value is $0.564\, \rm{nN \, \mu m}$ at power $630 {\rm mW}$ (corresponding to the output laser power of the experimental device). When using equation \ref{eq01} to calculate the $\tau_{max}$ it gives $3.35 \rm{\, nN \, \mu m}$ for a normalized torque value of $17\%$. \textcolor{black}{This value of torque is numerically calculated, but it depends on the real value of power reaching the OPTOBOT, the position of the laser beam, and the intensity of the Gaussian beam. An experimental approach would be recommended for more verification of this value.}.

\subsubsection*{Fabrication}
The microrobots were fabricated by a commercial $3$D printer (Photonic Professional GT+, Nanoscribe GmbH) utilizing the two-photon lithography technique. The photoresin chosen for the fabrication was the commercial IP-L $780$ negative resist (Nanoscribe GmbH) that can be used to print in oil immersion configuration over borosilicate D$263$ substrates with a thickness of $(170\pm10) \, {\rm \mu m}$. Moreover, IP-L $780$ provides high resolution where the smallest feature that can be printed using it is $200 \, {\rm nm}$.

 The printing was done layer-by-layer along the OPTOBOT's longest axis. The slicing and hatching distances were both equal to $0.1\, {\rm \mu m}$. A drop of the resist was deposited on the substrate and photopolymerized with a femtosecond laser operating at $\lambda= 780\, {\rm nm}$ and a $63$X objective in oil immersion configuration. A laser power of $50$ percent and a galvanometric scan speed of $4000\, {\rm \mu m/s}$ were used for the whole fabrication process. The robots were printed vertically to decrease the adhesion to the substrate, and a thin pillar with $660\, {\rm nm}$ diameter was printed under the helix to connect the OPTOBOTs to the substrate. Therefore, all the robots were supposed to be standing vertically over the pillar connected to the substrates.

After printing, the sample was developed for $12$ minutes in a propylene glycol methyl ether acetate (PGMEA) solution to remove all the unexposed photoresists.
The sample was further rinsed two times for $3$ minutes in isopropyl alcohol (IPA) to clear the developer. In order to dry these $3$D delicate microrobots efficiently, critical point drying was used (Autosamdri $931$, Tousimis).

\subsubsection*{SEM Imaging}
The fabricated structures were examined with an ApreoS ThermoFisher SEM in Optiplan mode. To achieve high-quality imaging, the surface was pre-coated with a thin layer of chromium using a Leica ACE$600 $sputter coater, achieving a chromium thickness of less than $5 {\rm nm}$. The imaging voltage was maintained below $10 {\rm kV}$ to prevent damage to the structures. 
The images are taken on theoretically identical structures used further for the optical experiments; however, they are not the same due to the metallic coating which is necessary for higher-resolution SEM images.

\subsubsection*{OPTOBOTs detachment from the substrate}
For the detachment of the microrobots from the substrate to be ready for optical manipulation, a platform centered around a commercial microscope (OPTIKA) is used. It includes a $3$-axis micromanipulator system (MP-$285$, Sutter Instrument). A micro-pipette (Borosilicate glass capillary, B$150$-$110$-$10$) was adjusted using a micropipette puller (Sutter Instruments, Model P-$1000$ Flaming Brown Micropipette Puller) to get a tip of diameter around $2 {\rm \mu m}$. The micro-pipette was attached to the micromanipulator system which was used to precisely control its movements to push the microrobots one by one and separate them from the substrate to be freely swimming in the solution. The solution is composed of water with $5\%$ of Tween $20$ to avoid adhesion of the robots to each other, or re-adhesion to the substrate (Supplementary Movie 1). The substrate used for the fabrication has a small thickness therefore when it is placed in the optical tweezers system, the objective lens can focus the laser beam over the surface of the substrate to reach the microrobots. Therefore, there was no need to transfer the microbots after their detachment from the fabrication substrate to a new substrate. The substrate used for the fabrication is used also for the optical manipulation after the robot's detachment.

\subsubsection*{Optical manipulation}

The optical tweezer platform shown in supplemental Figure S2 integrates an optical setup, a moving stage, and a controller device. The optical setup, built around a custom inverted microscope with an oil immersion objective (Olympus UPlanFLN $40$X, NA $1.3$), uses a near-infrared laser source ($1070 {\rm nm}$). Visual feedback is provided by LED illumination reflected in a high-speed CMOS camera (Basler, $659 \times 494$ px). The setup employs two actuation methods: a $3$D nano-stage (PI P-$562.3$CD) on a $2$D micro-stage (PI M-$126$.CGX) for large workspace and fine control, and high-speed laser deflection via a galvanometer (GVS$002$, Thorlabs) and a deformable mirror (PTT$111$ DM, Iris AO) for dynamic time-sharing multi-trap positioning (Supplementary Note 2). This setup allows for precise sample manipulation using the controller device that controls the movement of the micro and nanostage while keeping the optical traps stationary, combining passive and active actuation for improved control \cite{Edison}. 

After conducting the manipulation of the OPTOBOTs, the rotational velocity was determined by analyzing video recordings captured with the CMOS camera at a frame rate of $30 {\rm Hz}$. Individual frames were extracted using MATLAB and the helix's configuration was observed in each frame. A complete third rotation, attributed to the $C_3$ symmetry, was identified when the helix returned to the same configuration.

\section*{Supplementary Note 1: OPTOBOT geometry}
\begin{figure}[!h]
    \centering
     \includegraphics[width=12cm]{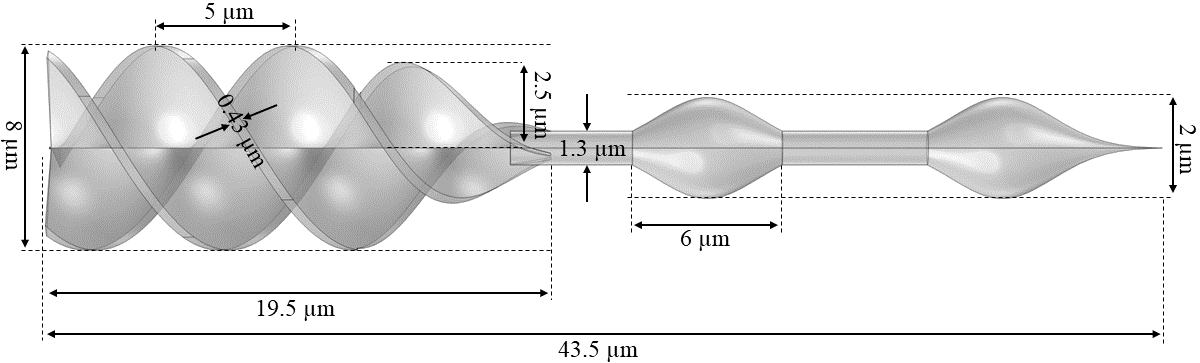}
    \caption{Side translucent view of the OPTOBOT showing the dimensions of its main parameters.}
    \label{geometry}
\end{figure}

Figure \ref{geometry}, shows the detailed geometry of the OPTOBOT presented in this study.

\section*{Supplementary Note 2: Optical setup}
\begin{figure}[!h]
    \centering
     \includegraphics[width=8.8cm]{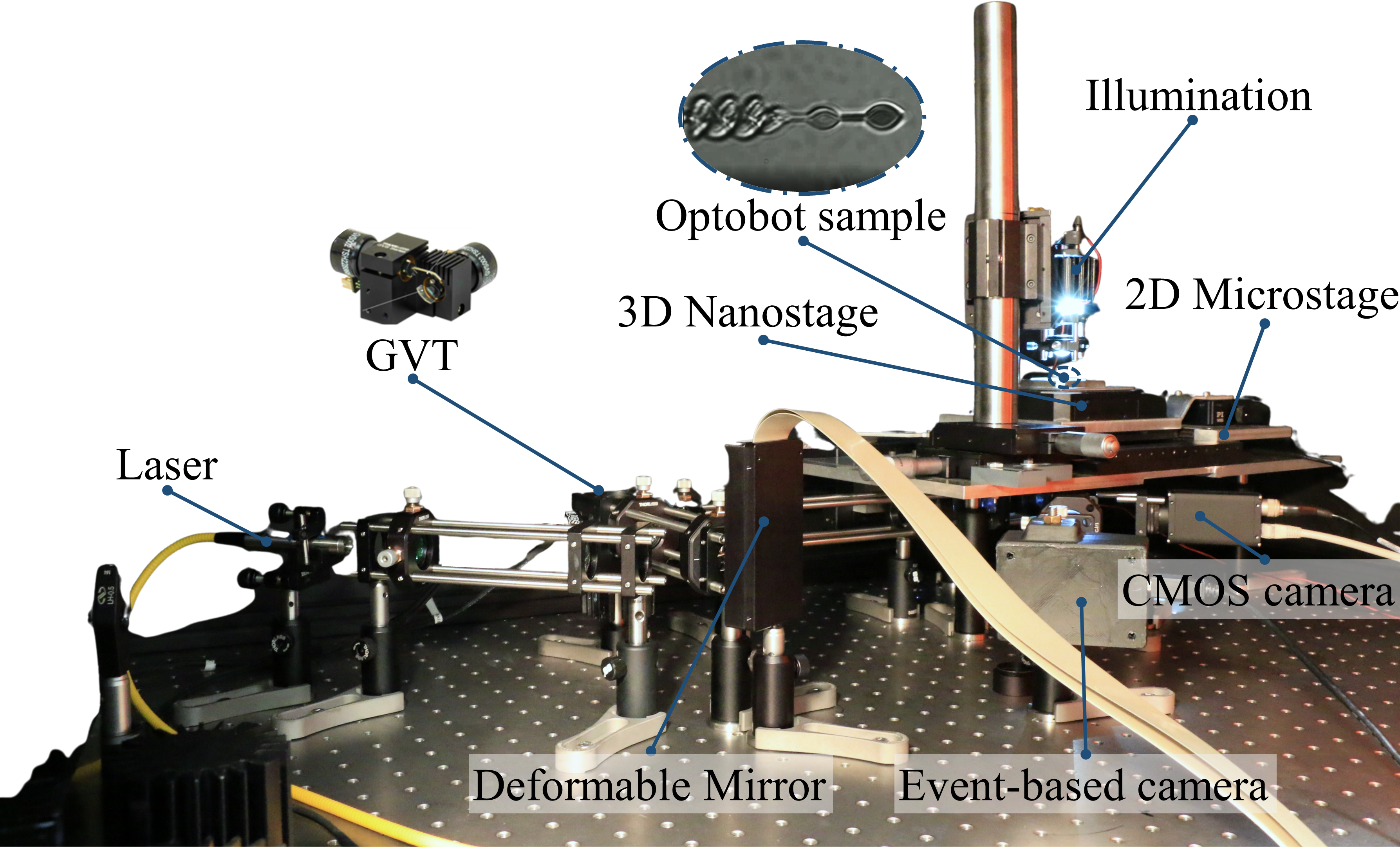}
    \caption{The optical tweezers system used in the experiments showing its components: the laser input, galvanometer, deformable mirror to give dynamic time-sharing multi-trap positioning, $3$D nano-stage on a $2$D micro-stage for position control, and LED illumination reflected in a high-speed CMOS camera for visual feedback.}
    \label{setup}
\end{figure}

Figure (\ref{setup}) shows the different components of the optical tweezer system used to produce all the data shown in the main paper. 

\section*{Supplementary Note 3: Influence of the size of the OPTOBOT}
\begin{figure*}[!h] 
    \centering
     \includegraphics[width=18cm]{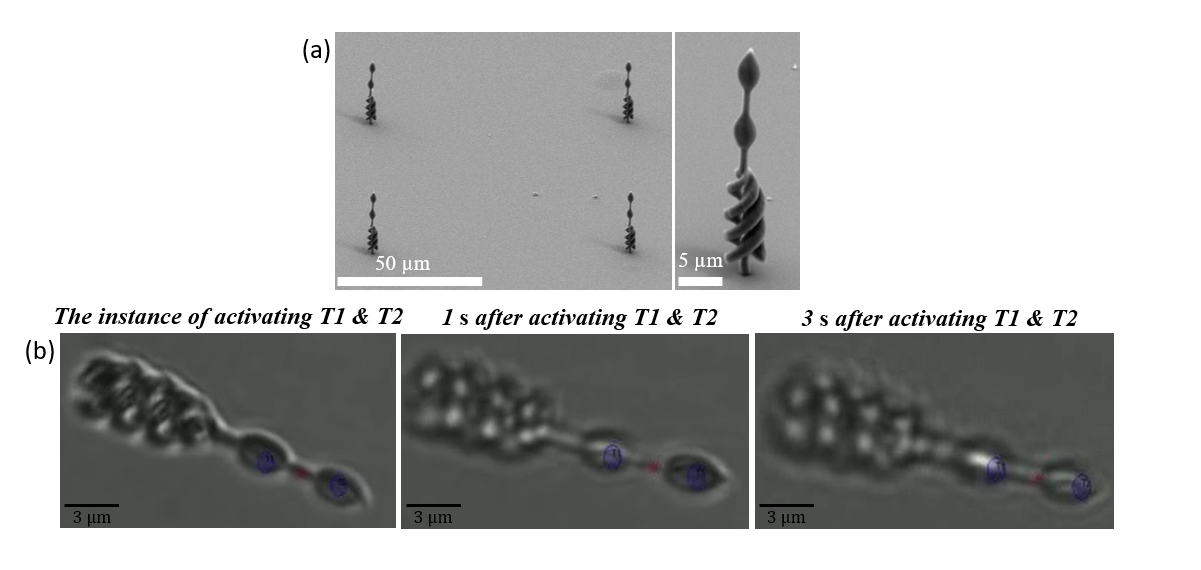}
    \caption{(a) SEM picture of the OPTOBOT with half the size. (b) When the laser traps with power 6.3\% are directed to the optical handles the robot is pushed and cannot be controlled. The snaphots are taken from video S6.}
    \label{size}
\end{figure*}

To study the effect of the size, another OPTOBOT with half of the size was fabricated (\autoref{size}(a)). However, trapping the small OPTOBOT was unsuccessful using power 6.3\%; even before adding the third trap on the helix, it was pushed and lost from the trap (see \autoref{size}(b) and video S6). As the size gets smaller, the distance between the optical handles and the helix becomes smaller. Thus, the interaction of the helix with the laser Gaussian beam increases leading to pushing the whole structure. The smaller size can be trapped with a lower power; however, the trap stiffness will be so low leading to unstable trapping of the OPTOBOT which can be easily separated from the optical trap by any external force.

\section*{Supplementary Note 4: Results reproducibility and power sweep}
\begin{figure*}[!h] 
    \centering
     \includegraphics[width=14cm]{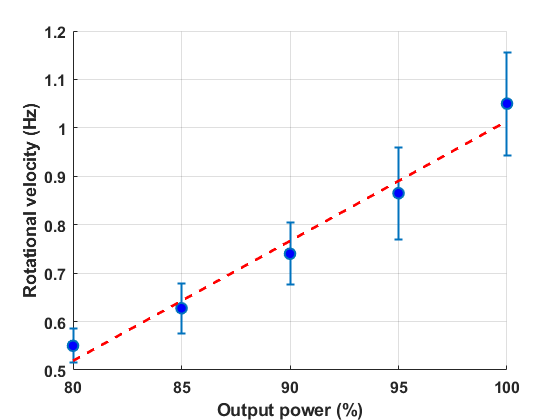}
   \caption{Results reproducing by a commercial optical tweezers system. The graph shows the dependence of rotational velocity on the laser power output from the laser source. \textcolor{black}{ A trend-line is drawn with error bars representing the standard error (SE).}}
    \label{ps}
\end{figure*}

To further validate the results, the OPTOBOT was tested using yet another commercial optical tweezer setup (Bruker JPK NanoTracker™ 2). The performance of the OPTOBOT was found to be comparable across both setups. Additionally, a laser power sweep was performed to determine the range of rotational velocities achievable by the OPTOBOT. The optical tweezers system used has a maximum output laser power of $5 \, \text{W}$; however, significant optical losses, approximately $90\%$, occur before the laser reaches the sample. Figure~\ref{ps} illustrates the rotational velocity as a function of laser power, expressed as a percentage of the device's maximum output. The results show that the rotational velocity is controlled by the laser power and can be increased up to $1.15\rm{Hz}$ (video S9).

\section*{Acknowledgements}

The authors acknowledge the support of ANR OPTOBOTs project (ANR-21-CE33-0003), ANR PNanoBot (ANR-21-CE33-0015) and Region Bourgogne Franche Comte project RobCell. The work was supported by the French RENATECH network and its FEMTO-ST technological facility MIMENTO and by the French research infrastructure ROBOTEX (TIRREX ANR-21-ESRE-0015) and its FEMTO-ST technological facility CMNR. This work has been achieved in the frame of the EIPHI graduate school (contract ”ANR-17-EURE-0002”).

\section*{Data availability}
The datasets used and/or analyzed during the current study are available from the corresponding author upon reasonable request. Data of figure 1(b) are included in the supplementary materials.

\bibliography{references}

\section*{Author contributions statement}
A.M.A. has done the design, numerical computations, experimental work, and writing the paper. 
A.M.A and E.G. have conducted together the experimental validations. 
A.M.A, J.A.I.M, and B.L. have contributed to the numerical aspect and the design of the OPTOBOTs.
G.U. has fabricated the samples.
A.M-O., S.H., A.B., and M.K. have contributed to the overall project supervision, discussions, and paper writing.
M.K. has contributed to the idea, the numerical simulations, and written the first draft of the paper.

\section*{Competing interests}
The authors declare no competing interests.
\section*{Materials and Correspondence}
Correspondence and requests for materials should be addressed to M.K. (email:muamer.kadic@femto-st.fr).

\end{document}